\documentclass[11pt]{article}

\usepackage[margin=1in]{geometry}
\usepackage{microtype}

\usepackage{graphicx}
\usepackage{amsmath,amssymb,amsfonts,amsthm}
\usepackage{mathtools}
\usepackage{booktabs}
\usepackage{multirow}
\usepackage{siunitx}
\usepackage{authblk}
\usepackage{xcolor}

\DeclareSIUnit{\GeV}{\giga\electronvolt}
\usepackage[numbers,sort&compress]{natbib}

\usepackage{hyperref}
\usepackage[nameinlink]{cleveref}

\title{Exploring the Boundaries of Differentiable Radiation Transport and Detector Simulation}

\author[1]{Jeffrey Krupa{$^{*}$}}
\author[2]{Yiyang Zhao}
\author[3]{Mihaly Novak}
\author[4]{Max Aehle}
\author[4]{Max Sagebaum}
\author[4]{Long Chen}
\author[4]{Nicolas Gauger}
\author[5]{David Lange}
\author[5]{Vassil Vassilev}
\author[6]{Miaoyuan Liu}
\author[7,8]{Lukas Heinrich}
\author[1]{Michael Kagan\footnote{Corresponding authors: \href{mailto:jkrupa@slac.stanford.edu,makagan@slac.stanford.edu}{jkrupa@slac.stanford.edu, makagan@slac.stanford.edu}}}

\affil[1]{SLAC National Accelerator Laboratory, Menlo Park, USA}
\affil[2]{Tsinghua University, Beijing, China}
\affil[3]{CERN, Geneva, Switzerland}
\affil[4]{RPTU University Kaiserslautern-Landau, Kaiserslautern, Germany}
\affil[5]{Princeton University, Princeton, USA}
\affil[6]{Purdue University, West Lafayette, USA}
\affil[7]{Technical University of Munich, Munich, Germany}
\affil[8]{Munich Center for Machine Learning (MCML), Munich, Germany}

\date{}

\begin{document}

\begingroup
\renewcommand{\thefootnote}{\arabic{footnote}}
\endgroup

\setcounter{footnote}{0} 
\renewcommand{\thefootnote}{\arabic{footnote}} 
\maketitle

\begin{abstract}

We present an application of automatic differentiation for particle transport through matter using a \textsc{Geant4}-like radiation transport simulation with a full electromagnetic physics model. 
When differentiating this step-based transport, we observe exploding gradients driven by rare but extreme sensitivities at material boundaries, which propagate through subsequent transport and shower development. 
To obtain usable derivatives for optimization, we introduce a targeted mitigation strategy that stops gradient propagation through boundary-crossing operations under identifiable unstable conditions while leaving the forward (primal) simulation unchanged.
We demonstrate that this enables stable, optimization-ready gradients in a detector-design problem.

\end{abstract}

\section{Introduction}\label{intro}

Radiation-transport simulations model the propagation of particles (such as photons, electrons, hadrons, etc.) through matter by combining deterministic particle navigation with stochastic sampling of physics interactions.
High energy physics (HEP) leverages the high-fidelity radiation transport framework \textsc{Geant4}~\cite{Agostinelli2003Geant4Toolkit,Allison2006Geant4Developments,Allison2016RecentGeant4} extensively to model energy flow inside massive detectors.
Outside of HEP, the same software is used extensively in applications including medical physics (e.g. DNA damage and radiation therapy~\cite{geant4-dna-0,geant4-dna-1,geant4-dna-2,geant4-dna-3,geant4-dna-4,geant4-dna-5}), space (e.g. instrumentation~\cite{geant-instrument-0}), and nuclear (e.g. shielding~\cite{geant-shielding-0,geant-shielding-1}).

High-energy physics detectors in particular have reached enormous complexity: modern collider experiments (e.g. the experiments at CERN's Large Hadron Collider) contain on the order of $10^8$--$10^{9}$ sensor elements. 
While these sensors are not tuned independently, their collective arrangement defines a hierarchically structured, high-dimensional design space spanning many design choices, such as geometry, materials, and operating conditions. 
Typically, the design of these detectors is informed by prior experience and hand-crafted heuristics, and tuned through resource-intensive simulation such as \textsc{Geant4}.
Informing the design of these high-dimensional systems using gradient-based optimization would have significant utility, if accurate derivatives can be obtained for observables with respect to key parameters like geometry and composition. 

More broadly, differentiable simulation---propagating derivatives through a physics simulator via automatic differentiation (AD)~\cite{baydin2018automaticdifferentiationmachinelearning}---has emerged as a powerful paradigm across machine learning and computational science.
By embedding a differentiable simulator inside a gradient-based optimization loop, one can solve inverse problems, tune design parameters, or train neural-network controllers end-to-end, without relying on surrogate models or finite-difference approximations.
This differentiable programming approach has been used in differentiable rendering~\cite{li2018diffmc,nimier2019mitsuba2}, robotics and mechanics~\cite{hu2020difftaichidifferentiableprogrammingphysical,newbury2024reviewdifferentiablesimulators}, and scientific computing~\cite{chen2019neuralordinarydifferentialequations}.
A recurring challenge in all these settings is that the underlying simulations contain discontinuities including visibility edges in rendering or boundary crossings in transport that can produce ill-defined or exploding gradients.

In this paper, we extend prior work~\cite{aehle2024} to present the first gradients obtained with full electromagnetic physics by applying AD to \textsc{hepemshow}~\cite{hepemshow-docs}, a lightweight transport application that implements the same stepping loop and physics engine (\textsc{g4hepem}~\cite{g4hepem-github}) used by \textsc{Geant4}, but with a simplified planar geometry.
In particular, we probe the accuracy, bias, and variance of derivatives by unpacking the stepper's inner workings, yielding important insights into differentiable radiation transport. 

Automating design with derivatives computed inside \textsc{Geant4} is a challenging technical task since the simulation itself is not built from the ground-up for propagating derivatives with AD. 
At its core, the simulation is non-smooth, relying on stochastic Monte Carlo (MC) sampling for modeling physics decays, and hard boundary decisions for modeling geometry. 
\textsc{Geant4} evolves showers by stepping individual particle tracks through volumes; at each step it samples interactions and may truncate the step at geometry boundaries. 
Indeed, as demonstrated in previous work~\cite{aehle2024}, naïvely differentiating through geometry navigation in \textsc{Geant4} leads to exploding derivatives that are ill-suited for gradient-based optimization.
For this reason, the physics engine used in these previous studies was incomplete.
Specifically, multiple Coulomb scattering (MSC) and random fluctuations were not simulated. 

We advance this prior work by extending it to the complete set of electromagnetic physics interactions, using MSC and fluctuations. 
To handle the exploding derivatives, we introduce a simple, physically-motivated mitigation rule that suppresses the dominant sources of gradient instability while leaving the forward simulation unchanged.
The resulting derivatives are stable and low-variance enough to drive gradient-based optimization, though they carry a controlled bias from the suppressed geometry terms.
We identify and characterize sources of gradient noise tied to geometry navigation, offering guidance for differentiable transport codes.

\begin{figure}
    \centering
    \includegraphics[width=0.64\linewidth]{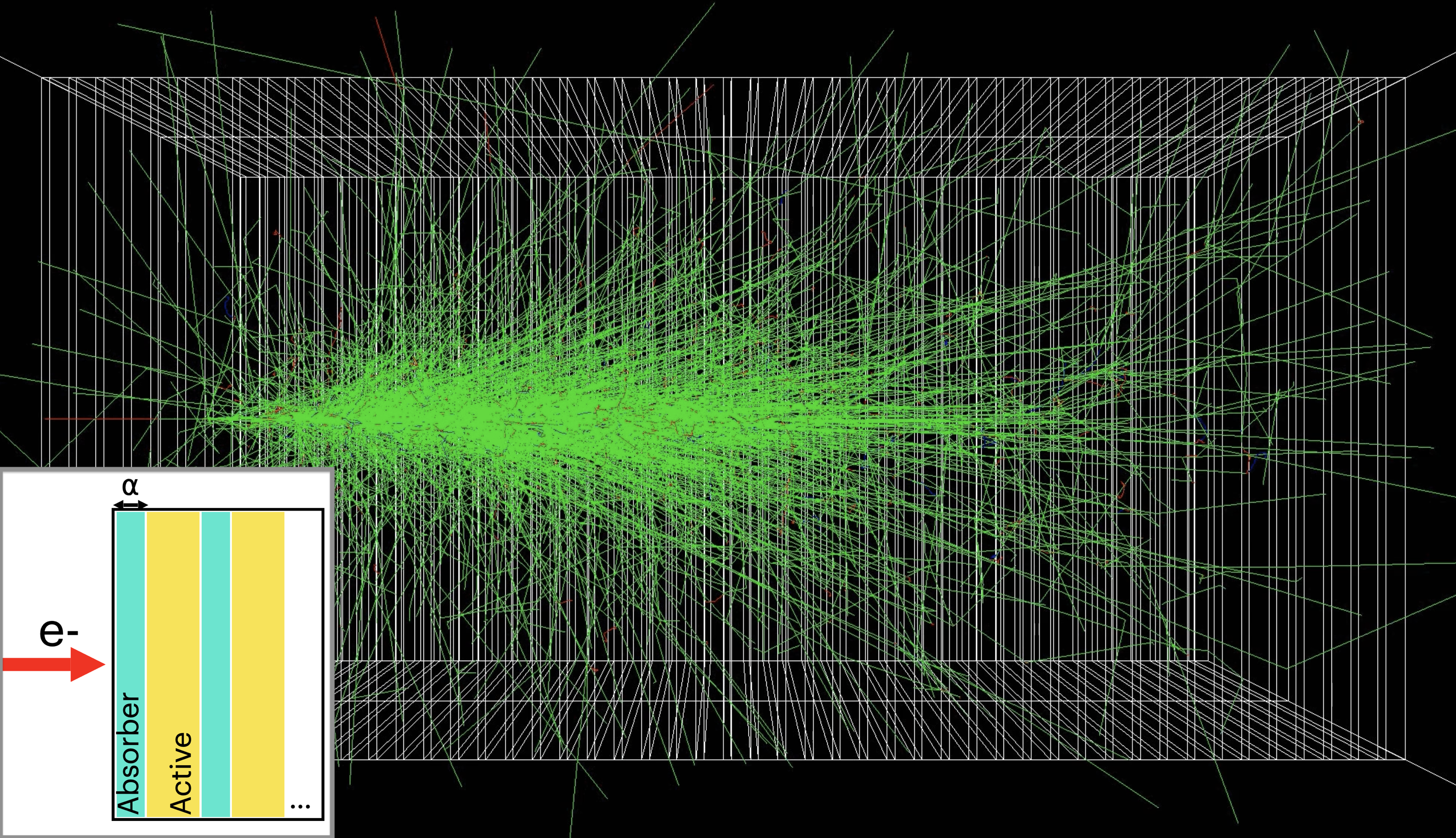}
    \includegraphics[width=0.35\linewidth]{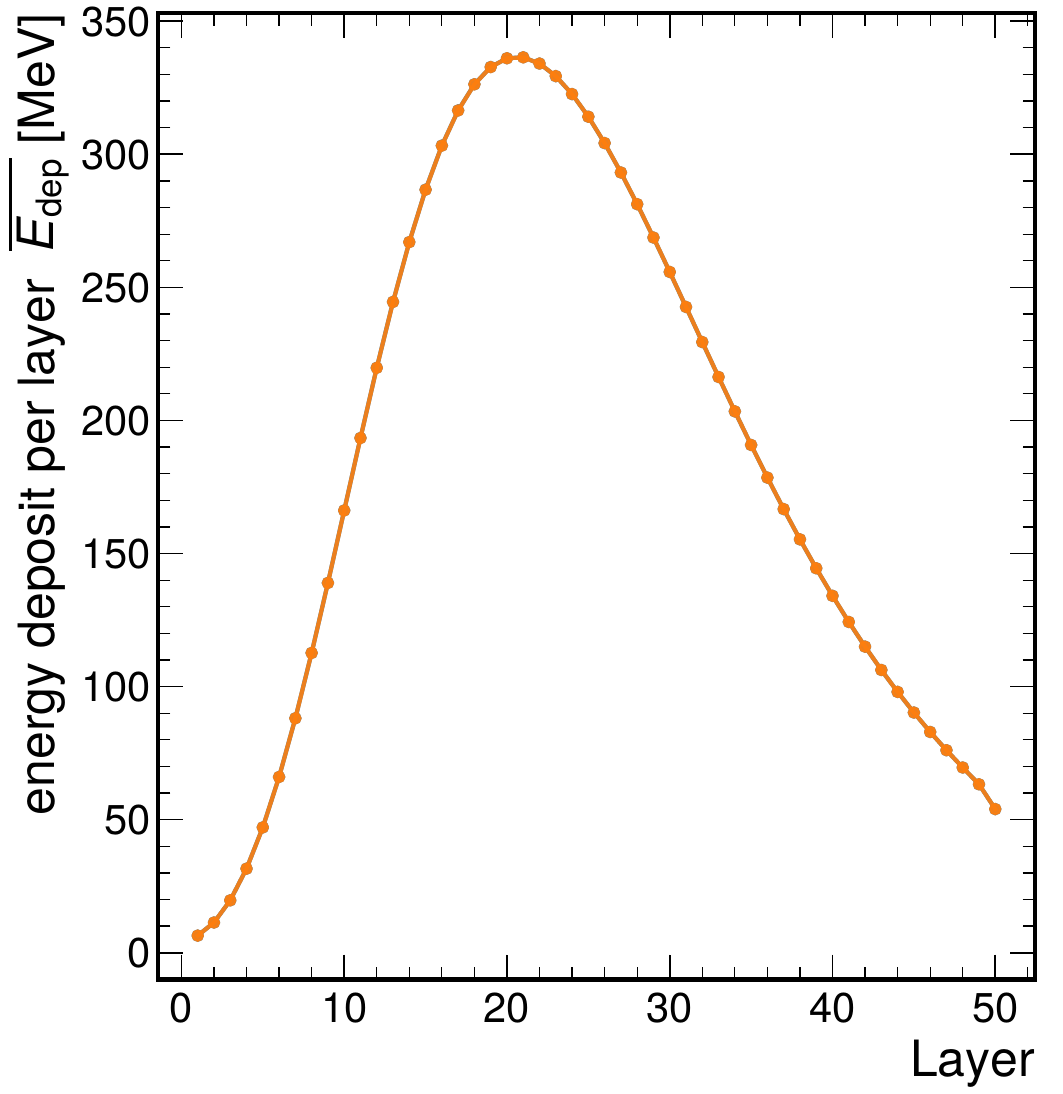}
    \caption{
    Left: Event display of a 50-layer sampling calorimeter illustrating an electromagnetic shower from an incident electron on the left side (inset: electron beam of energy $E_{\rm beam}$ entering a calorimeter comprised of alternating absorber/active layers with absorber thickness $\alpha$). 
    The complexity of the shower development motivates gradient-based optimization over manual tuning or grid search.
    Right: Mean longitudinal energy deposit per layer, showing the shower maximum in the middle layers.
    }
    \label{fig:event_display}
\end{figure}
As a case study, we consider a \emph{sampling calorimeter}, a device that measures particle energy through alternating layers of absorber and active material.
Calorimeters are central to radiation measurement across many domains: sampling and homogeneous calorimeters are used throughout collider and astroparticle physics~\cite{Fabjan:2003aq,fermi-lat} and water and graphite calorimeters serve as primary absorbed-dose standards in radiation therapy~\cite{Seuntjens2009}.
The layered geometry and well-understood physics of sampling calorimeters make them a canonical benchmark for detector optimization.

To demonstrate the complexity of this application, a single particle decay simulated by \textsc{Geant4} for a 50-layer sampling calorimeter comprised of lead (absorber layer) and liquid argon (active layer) is shown in~\autoref{fig:event_display}.
The incoming particle, a 10~GeV electron, produces up to $\mathcal{O}(10^4)$ distinct secondary particles, shown as green lines. 

In this paper, we contribute the following:

\begin{itemize}
	\item \textbf{Identification and treatment of instabilities in differentiable particle transport.} 
    We demonstrate that certain interactions with boundaries in stepping schemes can induce singular derivatives of step length with respect to geometry parameters. 
    These divergences propagate through the simulation chain and dominate event-level gradient variance.
    \item \textbf{A simple and physically consistent variance-control strategy.} 
    We introduce a local \emph{navigation-gradient stopping} rule that is applied only to geometry boundary navigation for tracks whose incidence angle with the local surface normal falls below a configurable threshold.
    This rule leaves the forward physics simulation unchanged while suppressing geometry-induced gradient noise by orders of magnitude.
    \item \textbf{Implications for differentiable detector simulation.} 
    Our findings provide practical criteria for numerical stability in differentiable transport codes and concrete guidance for applying geometry-aware AD in more complex settings.

\end{itemize}

The remainder of the paper is organized as follows.
We begin with a survey of related work on detector optimization and differentiable transport.
We then describe the stepping structure of radiation-transport simulations and identify the general mechanisms that produce exploding gradients at material boundaries, using a sampling calorimeter as a concrete case study.
Next, we present a variance-mitigation strategy based on selectively stopping gradients, which reduces variance at the cost of a small bias.
Finally, we demonstrate the resulting derivatives and their utility in an optimization task. 

\section{Related Work}\label{sec:related}

\subsection{Detector Design Optimization}
 
Traditional detector design in HEP relies on manual parameter tuning or grid searches over limited design spaces. 
Recent work has introduced a variety of data-driven approaches to systematically explore detector configurations. Several approaches focus on enabling gradient computations to drive gradient-based optimization, such as creating differentiable simulations with differentiable programming~\cite{aehle2024,aehle2024geant4,kagan2023branches,Cheong_2022,Strong:2023oew,Gasiorowski_2024,alterkait2026,AEHLE2025100120,Mokhtar_2025}, as well as surrogate approaches~\cite{Adelmann:2022ozp,Shirobokov:2020tjt, strong2024mutual,particles8020047} that train ML models to approximate simulators and enable differentiable optimization. The \textsc{MODE} collaboration is active in exploring end-to-end optimization of detectors with differentiable programming~\cite{DORIGO2023100085,Dorigo:2023kjv} and co-design of detectors and software~\cite{dorigo2026codesign}. Bayesian optimization has also proven effective for low-dimensional problems with expensive simulations~\cite{cerutti2020drich,ai-det-eic,PhysRevAccelBeams.27.084801}, offering sample-efficient black-box optimization without the need for gradient estimation. Additional approaches include evolutionary algorithms for multi-objective optimization~\cite{fanelli2023ecce} and reinforcement learning for sequential design decisions~\cite{qasim2024rl,Kortus:2025nav}. Recent work also explores how Large Language Models, agentic AI, and diffusion models can be used in detector design~\cite{LLMsforphysicsdesign,Nguyen:2026wsv,Hill:2026naa}.
Our work pursues direct differentiation of the underlying MC particle transport simulation.


\subsection{Differentiable Particle Transport}
 
Recent efforts have pursued multiple AD approaches directly on particle transport applications like \textsc{Geant4}. 
Recent work has integrated forward- and reverse-mode operator-overloading AD into \textsc{hepemshow}~\cite{aehle2024} and forward-mode operator-overloading AD into \textsc{Geant4}~\cite{aehle2024geant4}. 
Mean pathwise derivatives were computed for electromagnetic shower simulations in sampling calorimeters, achieving better performance than machine-code AD. 
A known limitation is the incompatibility of naïve AD application with MSC and fluctuations~\cite{aehle2024}.
 
These efforts have established the technical feasibility of differentiating particle transport codes and have validated derivatives on simple geometric configurations. 
However, prior work has not investigated gradient behavior in realistic optimization scenarios or addressed gradient pathologies that may emerge when gradients must guide actual design decisions. 
Our work identifies systematic sources of exploding gradients driven by rare but extreme sensitivities in geometry navigation and boundary-limited stepping, and introduces a targeted mitigation strategy that enables stable gradients for detector optimization problems.

\section{Background}\label{background}

\subsection{Electromagnetic transport and stepping}\label{sec:stepping}

Radiation-transport codes such as \textsc{Geant4} simulate the passage of particles through matter by tracking individual particles (``tracks'') through a sequence of discrete steps.
The electromagnetic (EM) physics model encompasses all of the key interactions that charged particles and photons undergo in matter: ionization, bremsstrahlung, pair production, annihilation, Compton scattering, the photoelectric effect, as well as MSC, which produces angular deflection of charged tracks.
Energy-loss fluctuations are also included.
Previous work~\cite{aehle2024} demonstrated that AD can be applied to \textsc{hepemshow}, but MSC and energy-loss fluctuations had to be disabled because they produced excessive gradient variance.
In this work, we include the complete set of EM interactions, including MSC and fluctuations, and address the resulting gradient instabilities.

At each step of the transport, physics imposes a constraint on the step length $L_{\mathrm{phys}}$ (e.g. due to discrete interactions, maximally allowed energy loss), while the geometry module computes the distance $L_{\mathrm{bnd}}$ to the nearest material boundary along the current direction.
The realized step length $L$ is the minimum of the two,
\begin{equation}\label{eq:step-min}
L=\min\!\left(L_{\mathrm{phys}},\,L_{\mathrm{bnd}}\right),
\end{equation}
so the update switches discontinuously between a physics-limited and a boundary-limited regime.
This ``min'' operation is the critical point where geometry enters the transport: when $L=L_{\mathrm{bnd}}$, small perturbations of geometry parameters (e.g.\ boundary position or material thickness) or of the track direction can induce large changes in $L$ and in the post-step state (position and material-region assignment).
These sensitivities propagate to downstream steps and ultimately to any observable accumulated along the particle shower.

\subsection{Gradient explosions at boundaries}\label{sec:observations}

\begin{figure}
    \centering
    \includegraphics[width=0.52\linewidth]{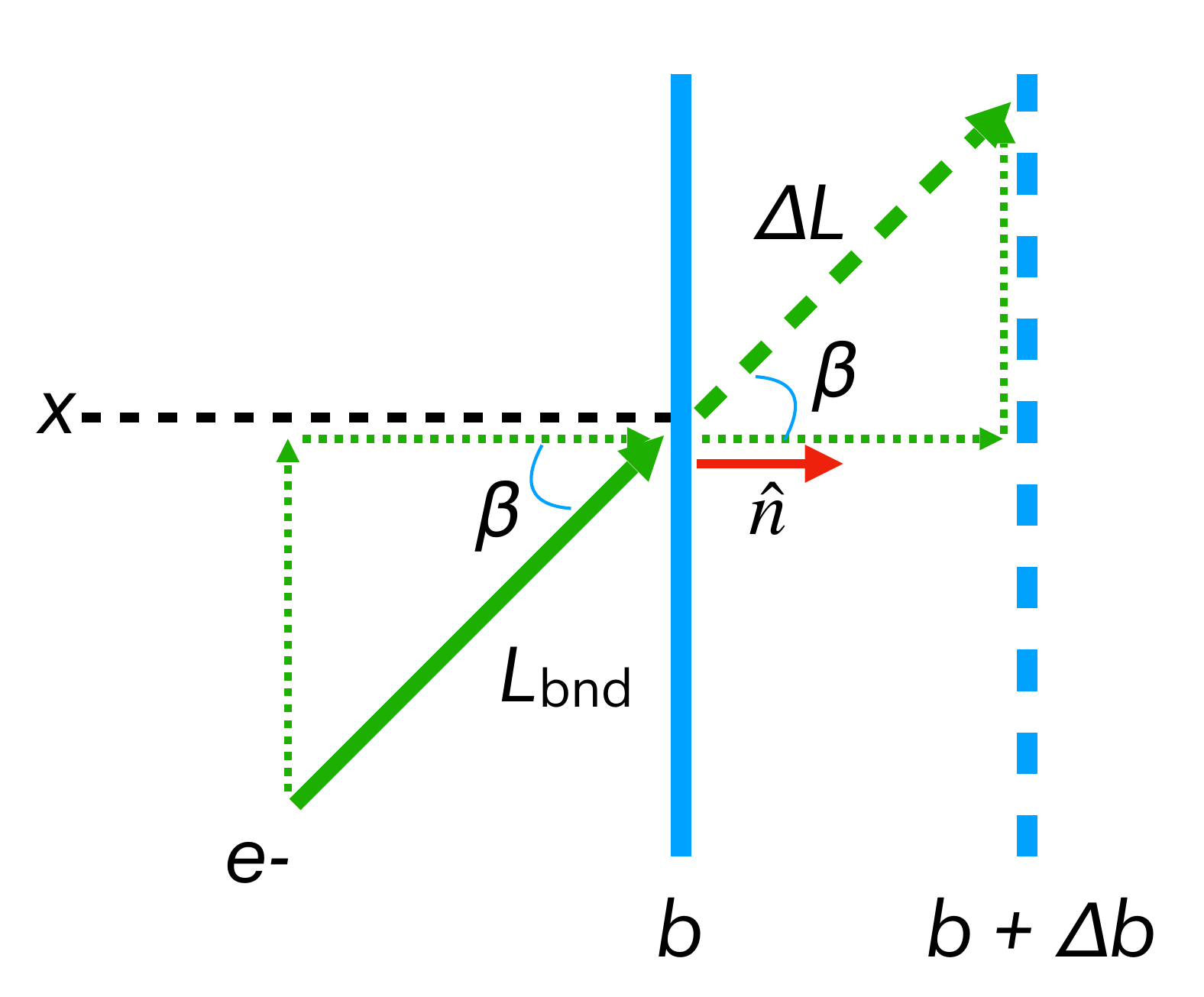}
    \includegraphics[width=0.46\linewidth]{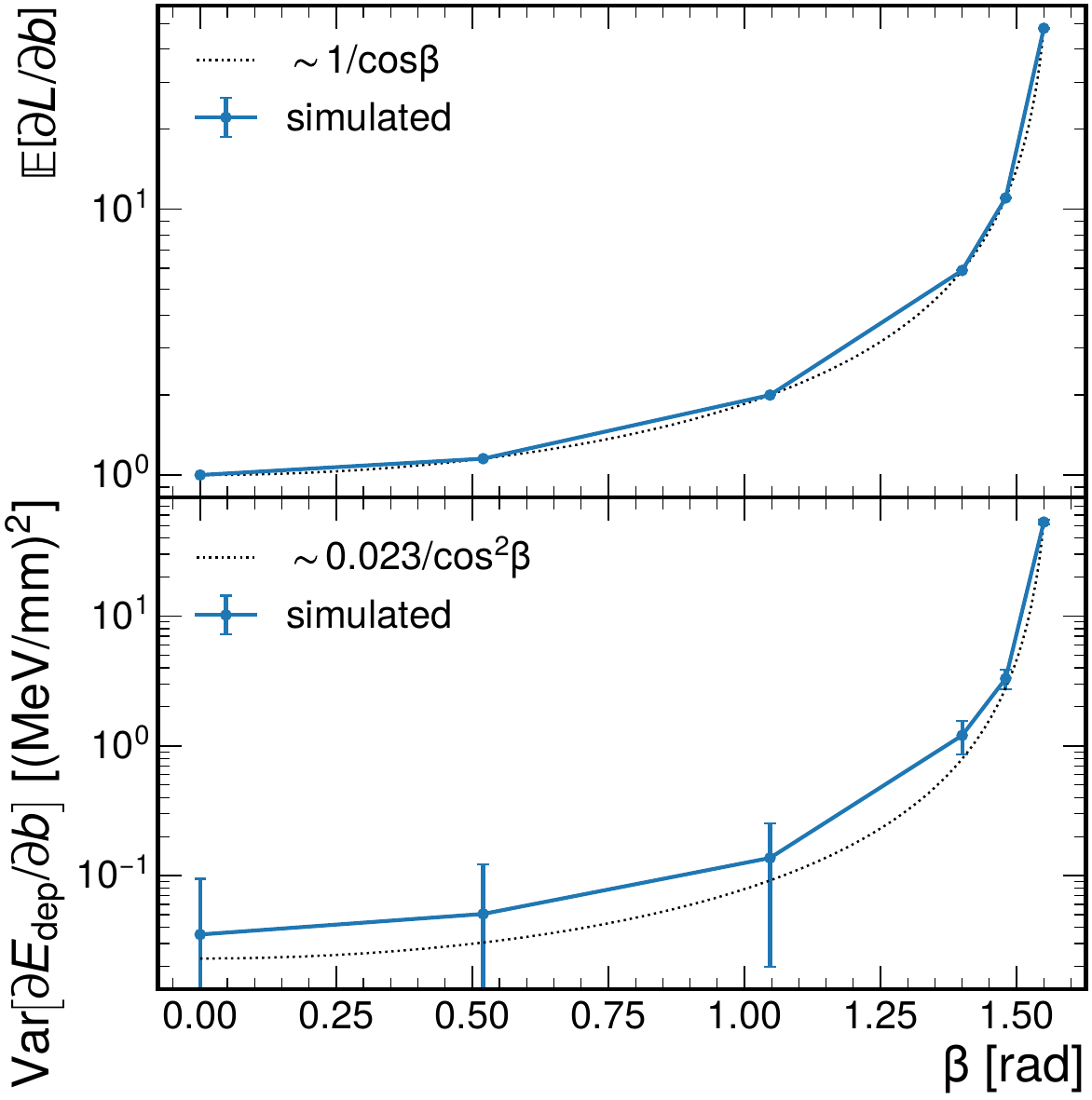}
    \caption{Left: a track of step length $L$ incident on a boundary with angle $\beta$ between the track direction $\hat{v}$ and the surface normal $\hat{n}$. Shifting the boundary by $\Delta b$ along $\hat{n}$ changes the step length by $\Delta L \propto \Delta b/\cos\beta$. Right: the mean derivative of the step length with respect to boundary position (upper) and the variance of the per-step energy-deposit derivative (lower), both as a function of incidence angle $\beta$, confirming the $1/\cos\beta$ divergence as $\beta\to\pi/2$.}
    \label{fig:what_is_happening}
\end{figure}

When a step is boundary-limited ($L=L_{\mathrm{bnd}}$), the realized step length depends directly on the intersection with the boundary.
In general, the sensitivity of the step length to a boundary displacement depends on the angle between the track direction $\hat{v}$ and the outward surface normal $\hat{n}$ at the point of intersection.
Defining $\cos\beta = \hat{v}\cdot\hat{n}$, a small shift $\Delta b$ of the boundary along $\hat{n}$ changes the intersection distance by
$\Delta L \approx \Delta b / \cos\beta$, implying
\begin{equation}\label{eq:slip}
\left|\frac{\partial L}{\partial b}\right| \propto \frac{1}{|\cos\beta|}.
\end{equation}
When a track travels nearly perpendicular to the normal vector ($|\cos\beta|\ll 1$, i.e.\ $\beta\to\pi/2$), this sensitivity diverges.
An illustration of this geometric ``slip'' is shown in a planar geometry on the left side of \autoref{fig:what_is_happening}: a small normal displacement of the boundary produces a large tangential displacement of the post-step point, which can alter subsequent boundary-distance queries and transport decisions.

For sufficiently small steps, and in the absence of branching or splitting, the continuous energy deposited over a step scales approximately with step length,
$E_\textrm{dep} \approx S(E)\,L$, where $S(E)$ is an energy-dependent stopping power.
Boundary-limited steps inherit the same angular amplification in $\partial E_{\mathrm{dep}}/\partial b$.
Consequently, the variance of the per-step energy-deposit sensitivities grows rapidly as $\beta\to\pi/2$, scaling parametrically as $\sim 1/\cos^2\beta$.

We validate this predicted scaling in a simplified two-layer geometry in which the beam position and boundary are arranged so that the first step is guaranteed to be boundary-limited.
The experiment is repeated at six angles between 0~rad and 1.55~rad with 10 events per angle.
For each event, we record the derivative at the first boundary-limited step only ($\partial L/\partial b$ and the corresponding $\partial E_{\rm dep}/\partial b$).
The measured mean of $\partial L/\partial b$ and the variance of $\partial E_{\textrm{dep}}/\partial b$ both diverge as $\beta\rightarrow\pi/2$, as shown in the right panels of~\autoref{fig:what_is_happening}.

In addition to near-parallel incidence, we observe a second failure mode that is strongly correlated with extreme gradient outliers: \emph{backward-going} boundary-limited steps.
Using per-track diagnostics recorded during transport, we classify tracks by whether they undergo at least one backward boundary-limited step (moving opposite to the nominal beam direction).
Although such steps are rare, they are disproportionately associated with exploding gradients.
In our simulations, only a small fraction of tracks (a few percent) exhibit backward boundary-limited behavior, and they contribute a similarly small fraction of the total energy deposit.
Nevertheless, these tracks account for more than 99\% of the total squared derivative ${\dot{E}_{\textrm{dep}}}^2$ accumulated over charged tracks, dominating the gradient variance.
This is consistent with the expectation that backward-going secondary particles at low energy undergo repeated boundary interactions, seeding large geometry-induced sensitivities that propagate through subsequent steps.

These two mechanisms---the $1/\cos\beta$ divergence at near-parallel incidence and the repeated-boundary instability for backward-going tracks---are general features of step-based transport through any geometry with material boundaries.
They are not specific to a particular detector type.

\subsection{Case study: sampling calorimeter}\label{sec:case-study}

To study these gradient pathologies concretely, we consider a 50-layer sampling calorimeter consisting of alternating absorber (lead) and active (liquid argon) layers, a canonical detector-design benchmark.
A mono-energetic electron beam enters normal to the layers, as shown in~\autoref{fig:event_display}, and the shower is simulated until all particle energy is deposited.

The parameters of interest are the incident beam energy $E_{\rm beam}$, absorber layer thickness $\alpha$, and active layer thickness $g$ (in this setup, each of the 50 absorber and active layers has the same thickness).
The key observable is the longitudinal energy-deposit profile: the total energy deposited in layer $i$ by all tracks, $E_{{\rm dep},i}$, averaged over events.
The collection of these per-layer deposits, denoted $E_\textrm{dep}$, is shown in~\autoref{fig:event_display} (right).
The longitudinal profile directly reflects the shower development, sampling structure, and material budget, making it a natural objective for detector optimization.
The baseline parameters are $E_{\rm beam} = 10,000$~MeV, $\alpha = 2.3$~mm, and $g = 5.7$~mm, which provide near-complete containment of the shower.

We denote sensitivities of the mean longitudinal energy-deposit profile by
$\partial \overline{E_{{\rm dep},i}}/\partial E_{\rm beam}$ and
$\partial \overline{E_{{\rm dep},i}}/\partial \alpha$.
In forward-mode AD these per-layer derivatives are referred to as dot-values (tangent-linear sensitivities), written generically as $\dot{E}_{{\rm dep},i}$.

Because all layer boundaries in this geometry are planar and perpendicular to the beam axis~$\hat{x}$, the outward surface normal is $\hat{n}=\pm\hat{x}$ and $\cos\beta = v_x$, the $x$-component of the track direction.
The near-parallel divergence of~\autoref{eq:slip} therefore corresponds to $|v_x|\ll 1$.

\begin{figure}[h!]
    \centering
    \includegraphics[width=.55\linewidth]{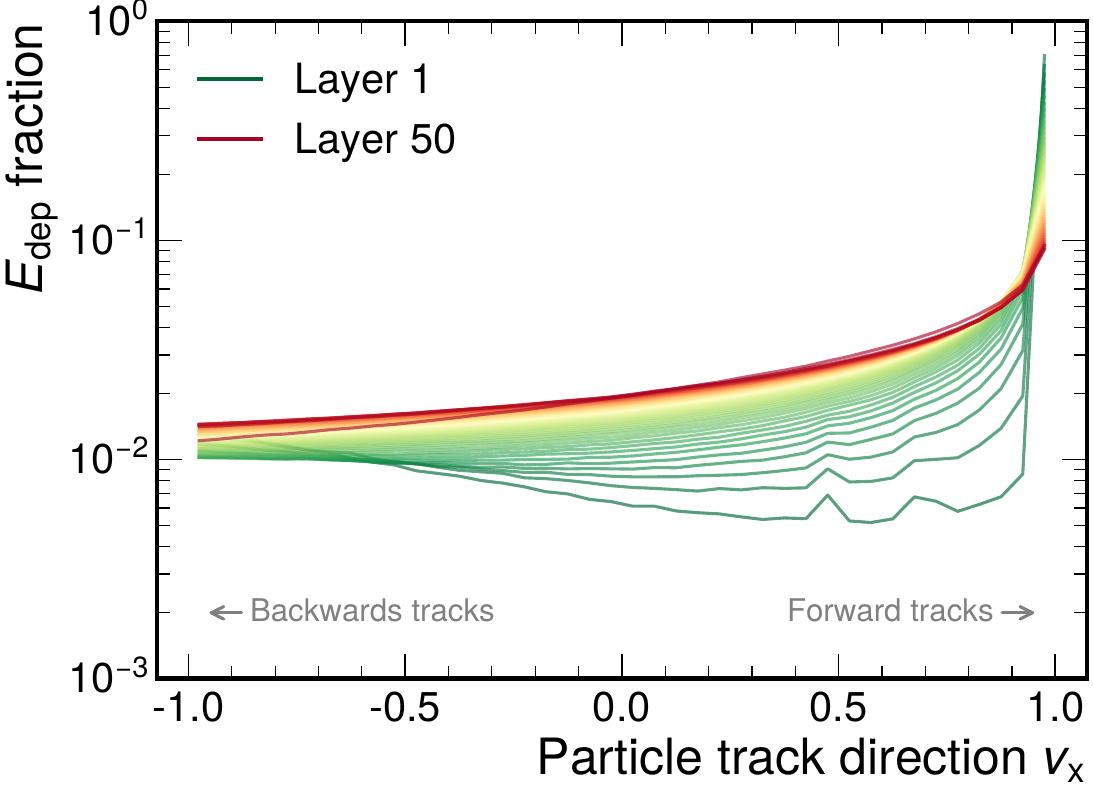}
    \caption{
    The fraction of track energy deposited as a function of the track direction projected onto the surface normal, $v_x = \hat{v}\cdot\hat{n}$.
    Each of the 50 layers is shown as a line in color gradients between green (layer 1) and red (layer 50).
    Later layers have a greater fraction of the deposit at small $|v_x|$ (near-parallel to the boundary) and $v_x < 0$ (backward-going), increasing the frequency of boundary-limited steps with large geometry-induced sensitivities.
    }
    \label{fig:efrac_direction}
\end{figure}

To illustrate how the general instabilities of~\autoref{sec:observations} manifest in this geometry, we examine how the deposited energy is distributed over track directions as a function of shower depth.
\autoref{fig:efrac_direction} shows the fraction of the charged-particle energy deposit as a function of $v_x =\hat{v}\cdot\hat{n}$.
Tracks moving parallel to the incident beam have $v_{\rm x}\approx +1$, while backward-going tracks have $v_{\rm x}\approx -1$.
Early layers are dominated by forward-going tracks, whereas later layers show an increasing contribution from large-angle and backward-going secondary particles, consistent with shower evolution toward lower energies and wider angular spread.
Both near-parallel incidence ($|v_{\rm x}|\ll 1$) and backward transport ($v_{\rm x}<0$) increase the probability of boundary-limited stepping and repeated boundary encounters, increasing exposure to the $1/\cos\beta$ divergence and the backward boundary-limited instability.

In the next section we describe a variance-mitigation strategy that selectively suppresses these geometry sensitivities while leaving the forward simulation and deposited-energy observables unchanged.

\section{Methods}\label{methods_and_related}

\subsection{Differentiable simulation and AD}
To explore these effects in a realistic particle simulator, we integrate an operator-overloaded AD package into a simplified framework with realistic electromagnetic physics simulation.
In particular, the operator-overloaded AD tool \textsc{CoDiPack}~\cite{codipack-github,SaAlGa2018OMS,SaAlGauTOMS2019} is integrated into a physics simulation toolkit, \textsc{hepemshow}~\cite{hepemshow-docs,hepemshow-github}. 
The methods here closely follow the original results presented in~\cite{aehle2024}.

Electromagnetic interactions are modeled using the \textsc{g4hepem} physics engine~\cite{g4hepem-github}, enabling a complete electromagnetic configuration in the studies presented here.
We perform sensitivity analysis using forward-mode AD to measure the change in the simulation output when simulation parameters are varied. 
Reverse-mode AD, which is also supported by \textsc{CoDiPack}, is used to obtain adjoints for the optimization studies.
The computational overhead introduced by forward and reverse mode AD is enumerated in~\cite{aehle2024}.

\subsection{Mitigating exploding gradients}\label{sec:mitigating-exploding-grads}

A naïve integration of AD produces derivatives with extremely large variance (“exploding gradients”).
As noted in~\cite{aehle2024}, these gradients are not useful for downstream optimization because their magnitudes can fluctuate by many orders of magnitude across otherwise similar events.
Motivated by the instabilities identified in~\autoref{sec:observations}, we introduce a simple strategy to mitigate exploding gradients that targets the dominant sources of instability.
The approach reduces gradient variance substantially at the cost of introducing a small bias.

Our core tool is a \emph{stop-gradient} operator (“stopgrad”), which blocks differentiation through a specific subset of the computation by setting the corresponding derivatives to zero.
Critically, the forward (primal) simulation is left unchanged, allowing consistency and reproducibility.
Such operators are widely used in machine-learning frameworks to prevent unstable or uninformative sensitivities from propagating.

In our implementation, we apply a stopgrad to the \emph{geometry-induced} sensitivities (position and direction updates) for boundary-limited steps that satisfy simple kinematic criteria based on the incidence angle $\cos\beta = \hat{v}\cdot\hat{n}$ between the track direction and the surface normal.
Specifically, we target:
\begin{enumerate}
    \item[(i)] \textbf{Near-parallel steps}: $|\hat{v}\cdot\hat{n}| < f$, where the threshold $f$ controls how close to grazing incidence a track must be before triggering the stopgrad. These are steps for which the $1/\cos\beta$ sensitivity amplification identified in~\autoref{sec:observations} is largest.
    \item[(ii)] \textbf{Backward-going steps}: $\hat{v}\cdot\hat{n} < 0$, which are empirically associated with repeated boundary interactions and large accumulated sensitivities (cf.~\autoref{sec:observations}).
\end{enumerate} 
In the planar geometry of our sampling calorimeter, $\hat{n} = \pm\hat{x}$ and $\hat{v}\cdot\hat{n} = v_x$, so these conditions reduce to $|v_x|<f$ and $v_x < 0$, respectively.
More generally, for non-planar geometries, the local surface normal at each boundary intersection would be used.

Once a step triggers the stopgrad, the geometry-induced sensitivities for all subsequent steps of that track are also suppressed, since those steps inherit the perturbed post-step state and would otherwise reintroduce the same instability.
A further choice concerns daughter particles created by the flagged track: their initial state also inherits the unstable sensitivities.
We therefore consider two propagation rules and compare their effect:
\begin{itemize}
    \item[(i)] \textbf{Track-only}: stopgrad is applied to the remainder of the flagged track; daughter particles are \emph{not} affected.
    \item[(ii)] \textbf{Track + descendants}: stopgrad is applied to the remainder of the flagged track \emph{and} to all particles it subsequently produces.
\end{itemize}
The track + descendants rule suppresses more derivative contributions and is used as the default configuration in this work (see~\autoref{tab:stopgrad_fractions} for a comparison). 
Further studies showing how the derivatives depend on $f$ are presented in~\autoref{sec:bias-analysis}. 
Note that choosing a parameter $f$ depends on the bias and variance requirements of the application being considered.

\begin{figure}[h!]
    \centering
    \includegraphics[width=0.99\linewidth]{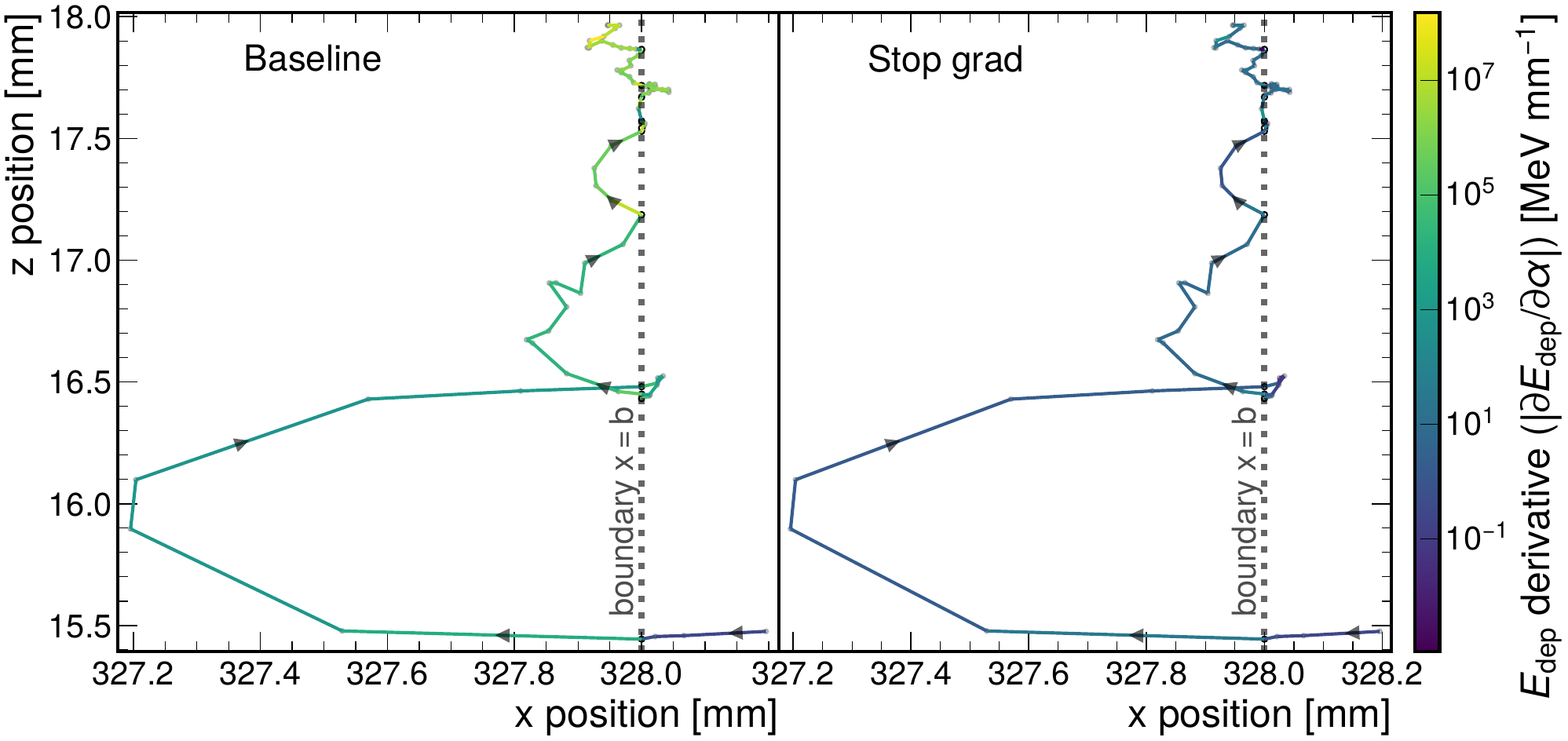}
    \caption{
        An example of a low-energy track exhibiting repeated boundary-limited steps near boundary $x=b$ in layer 41. 
        Baseline AD (left) produces extreme $|\partial E_{\textrm{dep}}/\partial \alpha|$ spikes during backscatter/ping-pong near the interface, with values accumulating  to $10^{8} \;\rm MeV mm^{-1}$ over 86 steps. 
        Applying the geometric stopgrad for backward/near-parallel (right) steps suppresses these unstable geometry sensitivities by about $10^{6}$. 
        The primal mode and trajectory remain unchanged.
    }
    \label{fig:single_track_display}
\end{figure}

The effect of the stopgrad is illustrated in \autoref{fig:single_track_display}, which shows a representative low-energy charged track in the baseline AD configuration (left) and with our stopgrad protection enabled (right). 
The track originates near the bottom of the panel (with kinetic energy 1.6~MeV) and propagates in the negative-x direction toward the interface at $x=b$ (arrows indicate the direction of transport).
Initially, the magnitude of the local sensitivity $|\partial E_{\rm dep}/\partial \alpha|$ is modest, $\mathcal{O}(10^{-1})$.
After several large-angle deflections from MSC---a common feature at low energies---the track repeatedly encounters the boundary and undergoes a sequence of boundary-limited steps near the interface.
In the unprotected configuration, these repeated boundary interactions lead to large, intermittent spikes in $|\partial E_{\rm dep}/\partial \alpha|$ that accumulate over many steps, reaching $\mathcal{O}(10^{8})\ \mathrm{MeV\ mm^{-1}}$ by termination.
With the stopgrad enabled, the primal trajectory is unchanged, but the unstable geometry-induced sensitivities are suppressed and the maximum derivative remains at $\mathcal{O}(10^{2})\ \mathrm{MeV\ mm^{-1}}$.

Lastly, two additional \emph{derivative-only regularizations} address near-singular Jacobian terms that arise in specific physics sub-models.
In both cases we apply the same primal-preserving pattern: for a function $f(\mathbf{x})$ whose exact Jacobian $\nabla f$ may diverge when an intermediate quantity approaches zero, we write the regularized output as
\begin{equation}\label{eq:keep-primal}
\tilde{f}(\mathbf{x}) \;=\; \underbrace{\mathrm{sg}\!\bigl(f(\mathbf{x})\bigr)}_{\text{exact primal}}
\;+\; \underbrace{\nabla f_{\mathrm{reg}}\!\bigl(\mathrm{sg}(\mathbf{x})\bigr) \cdot \bigl(\mathbf{x} - \mathrm{sg}(\mathbf{x})\bigr)}_{\text{regularized first-order correction}},
\end{equation}
where $\mathrm{sg}(\cdot)$ denotes the stop-gradient operator (identity on values, zero on derivatives) and $\nabla f_{\mathrm{reg}}$ is the Jacobian of $f$ evaluated with near-singular denominators or arguments floored/capped at a threshold.
By construction, $\tilde{f}$ returns the exact forward value $f(\mathbf{x})$ but injects a well-behaved Jacobian into the AD tape.

\paragraph{MSC path-length conversion.}
The MSC model is responsible for accounting for the net effects of the large number of small angular deflections along the condensed history step of charged particles as they traverse matter.
The simulation therefore distinguishes between the \emph{true path length}, $t$, and the \emph{geometrical path length}, $z$, of a condensed history step. The former corresponds to the length of the actual path traversed between the pre- and post-step points if the many angular deflections are included, while the latter is the length of the projection of the vector pointing from the pre- to the post-step point along the original direction.
In the MSC model used here~\cite{Urban2006}, the two are related in the short-step regime (i.e. when energy loss along the step can be neglected) by
\begin{equation}\label{eq:mcs-conv}
  z \;=\; \lambda_1\!\left(1 - e^{-t/\lambda_1}\right),
\end{equation}
where $\lambda_1$ is the first transport mean free path for elastic scattering.
When a boundary-limited step shortens the geometric path, the simulator must invert this relation to recover the true step length:
\begin{equation}\label{eq:mcs-inv}
  t \;=\; -\lambda_1\,\ln\!\Bigl(1 - \frac{z}{\lambda_1}\Bigr),
  \qquad
  \frac{\mathrm{d}t}{\mathrm{d}z} \;=\; \frac{1}{1 - z/\lambda_1}\,.
\end{equation}
The derivative diverges as $z\to\lambda_1$.
More generally, when energy loss along the step is accounted for, the conversion formula introduces additional rational functions of~$\lambda_1$ in the denominator, all of which can become near-singular when the step length approaches the particle's remaining range.
We regularize by evaluating the Jacobian with all such denominators floored at a threshold~$\epsilon_{\mathrm{conv}}$, following~\autoref{eq:keep-primal}.

\paragraph{Photon step limit.}
The step length of a photon until the next physics interaction (Compton scattering, pair production, photoelectric absorption) follows an exponential distribution with the inverse macroscopic cross section, called the mean free path (MFP), as its parameter. Therefore, the step length until the next physics interaction is sampled by drawing the number of mean free paths to the next interaction, $n_{\mathrm{left}} = -\ln (u)$ with $u \sim \mathrm{Uniform}(0,1)$, and converting it to a step length by $L_{phys} = \lambda n_{\mathrm{left}}$ using the actual value of $\lambda(E,\mathrm{Material})$, the energy- and material-dependent MFP.
After each step of length~$\ell$ and MFP of $\lambda$, the counter is decremented by removing the corresponding number of mean free paths:
\begin{equation}\label{eq:numia-update}
  n_{\mathrm{left}} \;\leftarrow\; n_{\mathrm{left}} - \frac{\ell}{\lambda}\,.
\end{equation}
When $\lambda$ is very large (i.e.\ the material is nearly transparent at a given photon energy), the step limit $L_{phys} = \lambda\,n_{\mathrm{left}}$ amplifies any sensitivity in~$\lambda$ into a correspondingly large sensitivity in the proposed step length.
Following~\autoref{eq:keep-primal}, we regularize by capping the MFP at $\lambda_{\mathrm{clip}}$ in the derivative evaluation.

Scans over both $\epsilon_{\mathrm{conv}}$ and $\lambda_{\mathrm{cap}}$ are shown in~\autoref{sec:appendix}.

\subsection{Experimental setup}

In addition to computing derivatives of the \textsc{hepemshow} simulation using AD with \textsc{CoDiPack}, we calculate derivatives with finite differences, providing independent validation of the derivatives. 
A central difference is computed for the energy deposited in each layer, $E_{\textrm{dep},\ell}$, with respect to a parameter.
A common set of random numbers is used to seed the forward and backward differences. 
While finite differences serve as a useful cross-check, they become impractical in realistic settings: each additional parameter requires two extra simulation passes, making the cost scale linearly with the number of parameters; and achieving low-variance estimates requires large event samples for each evaluation, compounding the expense.
AD, by contrast, obtains all derivatives in a single simulation pass (forward mode per parameter, or reverse mode for all parameters simultaneously), making it the only viable route for high-dimensional or large-sample sensitivity analyses.

To demonstrate the usefulness of these derivatives in an optimization setting, we apply the derivatives with AD in the context of a two-parameter optimization, closely following the method in~\cite{aehle2024}.
First, a reference longitudinal energy-deposit profile is computed at the nominal parameters $(E_{\mathrm{beam}}^*,\alpha^*) = (10^4~\mathrm{MeV},\,2.3~\mathrm{mm})$. 
The energy-deposit profile is estimated using 100k events. 
The optimization loop then proceeds by minimizing the mean-squared-error between the current longitudinal shower profile and the reference profile:
\begin{equation}
L(E_{\mathrm{beam}},\alpha)=\sum_{i=1}^{50}\left(\overline{E_{\mathrm{dep}}}_i(E_{\mathrm{beam}},\alpha)-\overline{E_{\mathrm{dep}}}_i(E_{\mathrm{beam}}^*,\alpha^*)\right)^2 .
\end{equation}
At each iteration of the optimization (epoch), we estimate $\overline{E_{\mathrm{dep}}}_i(E_{\mathrm{beam}},\alpha)$ using 5k events, and compute gradients with the AD-enabled simulation.
Our variance-mitigation rules are applied. 

We use a mini-batch stochastic gradient descent (SGD) approach for the optimization. 
The parameter updates are clipped to stabilize occasional large steps.
The initial clipping parameters are $(
\Delta E_{\mathrm{beam}}^{\textrm{max}},\Delta \alpha^{\textrm{max}}) = (2000~\mathrm{MeV},\,1~\mathrm{mm})$.
The learning rate is chosen such that typical parameter updates are comparable in scale across the two coordinates.
The learning rate and clipping parameters are reduced by a factor of 2 after 30 epochs with no improvement in the best loss.
The optimization terminates after 3 decays. A maximum of 200 epochs is considered.

\section{Results}\label{sec2}

\subsection{Gradients}
\begin{figure}[h!]
    \centering
    \includegraphics[width=0.49\linewidth]{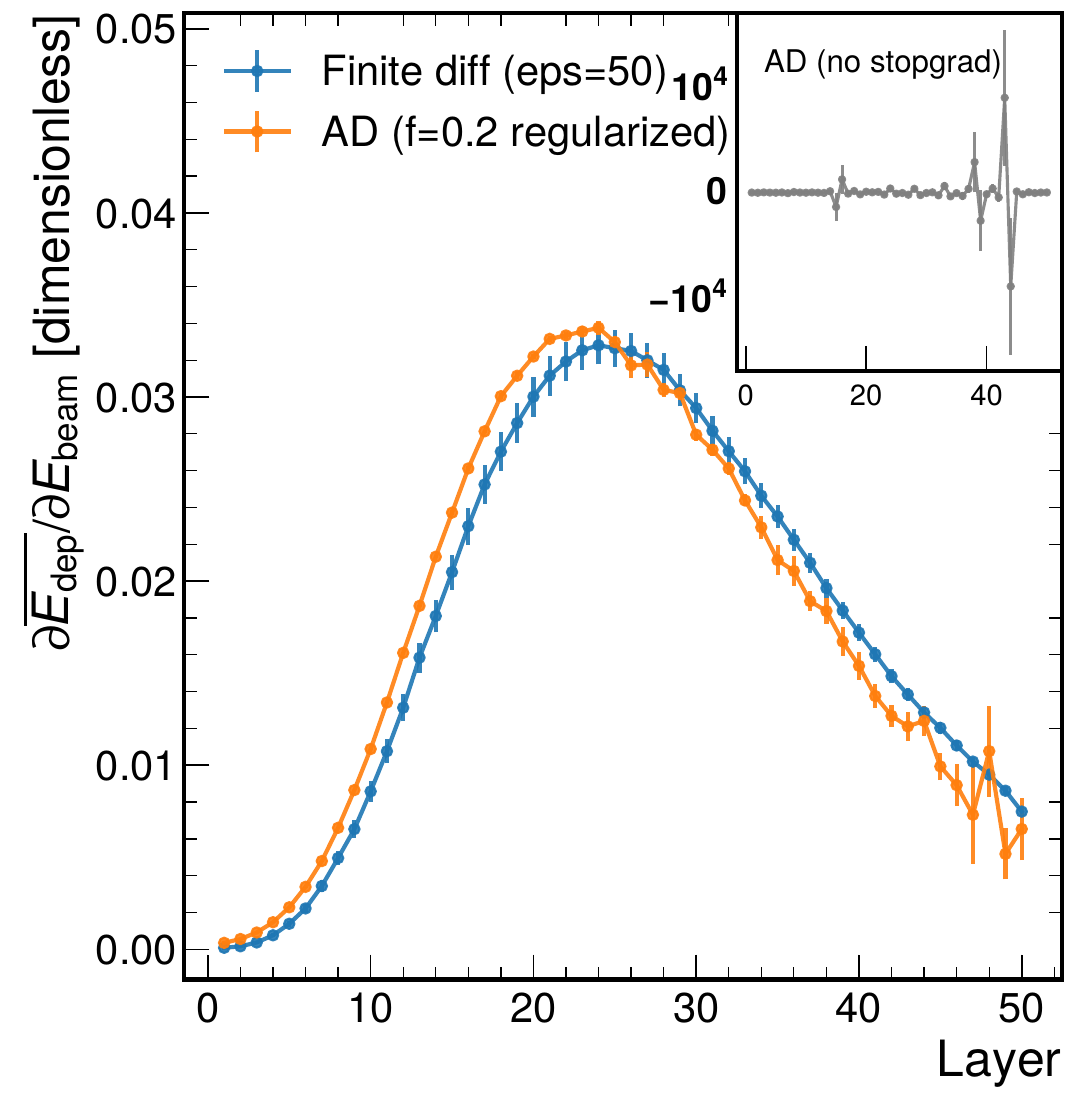}
    \includegraphics[width=0.49\linewidth]{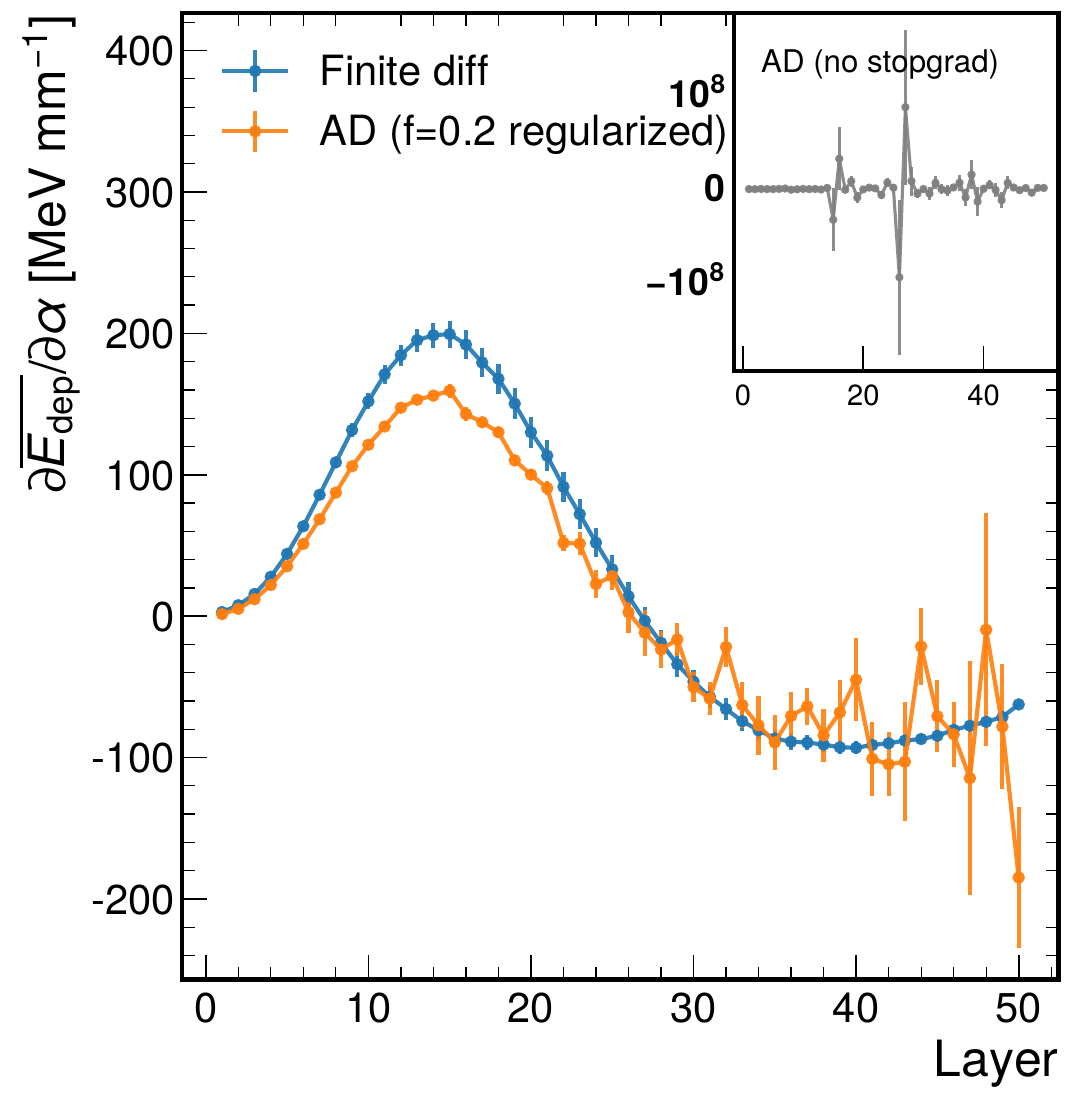}  

    \caption{
    Mean longitudinal energy-deposit derivative per calorimeter layer comparing finite differences (blue) to AD (orange) with the navigation stop-gradient rule applied at boundary-limited steps with $\hat{v}\cdot\hat{n}<0.2$ ($v_x<0.2$).
    Left: derivative with respect to the beam energy, $\partial E_\mathrm{dep}/\partial E_{\mathrm{beam}}$, showing close agreement in shape and scale between AD and finite differences across layers. 
    Right: derivative with respect to absorber thickness, $\partial E_\mathrm{dep}/\partial \alpha$, where the stopgrad rule yields a stable, finite-variance estimate.
    The insets show the unprotected AD estimator (``no stopgrad'') exhibiting rare, catastrophic outliers and orders-of-magnitude larger fluctuations (note the difference in scales). 
    Error bars show $\sigma/\sqrt{N}$ per layer.
    }
    \label{fig:derivatives}
\end{figure}

Forward-mode derivatives from \textsc{hepemshow} after applying our variance-mitigation (stopgrad) rule are shown in \autoref{fig:derivatives}. 
The left and right panels show $\partial \overline{E_{\mathrm{dep}}} / \partial E_{\mathrm{beam}}$ and $\partial \overline{E_{\mathrm{dep}}} / \partial \alpha$, respectively, as functions of calorimeter layer. 
The AD result (orange) applies the stopgrad to geometry-induced sensitivities on boundary-limited steps with $\hat{v}\cdot\hat{n}<0.2$ (i.e.\ $v_x<0.2$ in our planar geometry), covering all backward-going steps and near-parallel steps limited by a boundary. 
Derivatives are estimated from $10^{8}$ events; error bars show the standard error on the mean ($\sigma/\sqrt{N}$) per layer.
Finite-difference (blue) derivatives use a central difference with step sizes $h=50$~MeV for $E_{\mathrm{beam}}$ and $h=0.005$~mm for $\alpha$ (with common random numbers between $p\pm h$)\footnote{Scans against step size show that derivatives are insensitive to choice of $h$.}. 
For reference, the unprotected AD estimator is shown in an inset with a different y-scale.

Without protection from the stopgrad, the AD estimator is dominated by rare outliers, producing layer-wise fluctuations that are 6 orders of magnitude larger than the stopgrad result.
This prevents comparison on a single axis; a different y-axis is used in the inset (no stopgrad).
The variance-mitigated AD results agree with finite differences within statistical uncertainty for $E_{\mathrm{beam}}$.
For $\alpha$, the overall shape and peak location are consistent with finite differences, but AD under-estimates the peak magnitude by approximately 25\%, indicating a residual bias introduced by suppressing geometry-driven sensitivities in the central region. 

Uncertainties increase in the late layers ($\gtrsim 35$) since the deposited energy per event is small and the shower becomes more angularly diffused: a larger fraction of the energy deposit comes from large-angle and backward-going secondary particles (cf.~\autoref{fig:efrac_direction}). 
In this regime, low-energy charged tracks undergo stronger MSC deflections and more frequent boundary interactions, leading to a heavier-tailed per-event derivative distribution and hence larger $\sigma/\sqrt{N}$. 
While the stopgrad rule suppresses catastrophic outliers by $\sim 6$ orders of magnitude, additional variance-reduction strategies may further stabilize the late-layer estimates.
Overall, these derivatives are stable and well-suited for downstream optimization, as shown in~\autoref{sec:optimization}.

\subsection{Bias analysis}\label{sec:bias-analysis}

To understand the bias in the derivative caused by the derivative regularization methods introduced in
~\autoref{sec:mitigating-exploding-grads},
we measure the energy deposit derivative with respect to absorber thickness by scanning each regularization parameter independently while holding the others fixed.
The resulting AD derivatives are then compared against
central finite differences.

\begin{figure}[h!]
    \centering
    \includegraphics[width=0.48\linewidth]{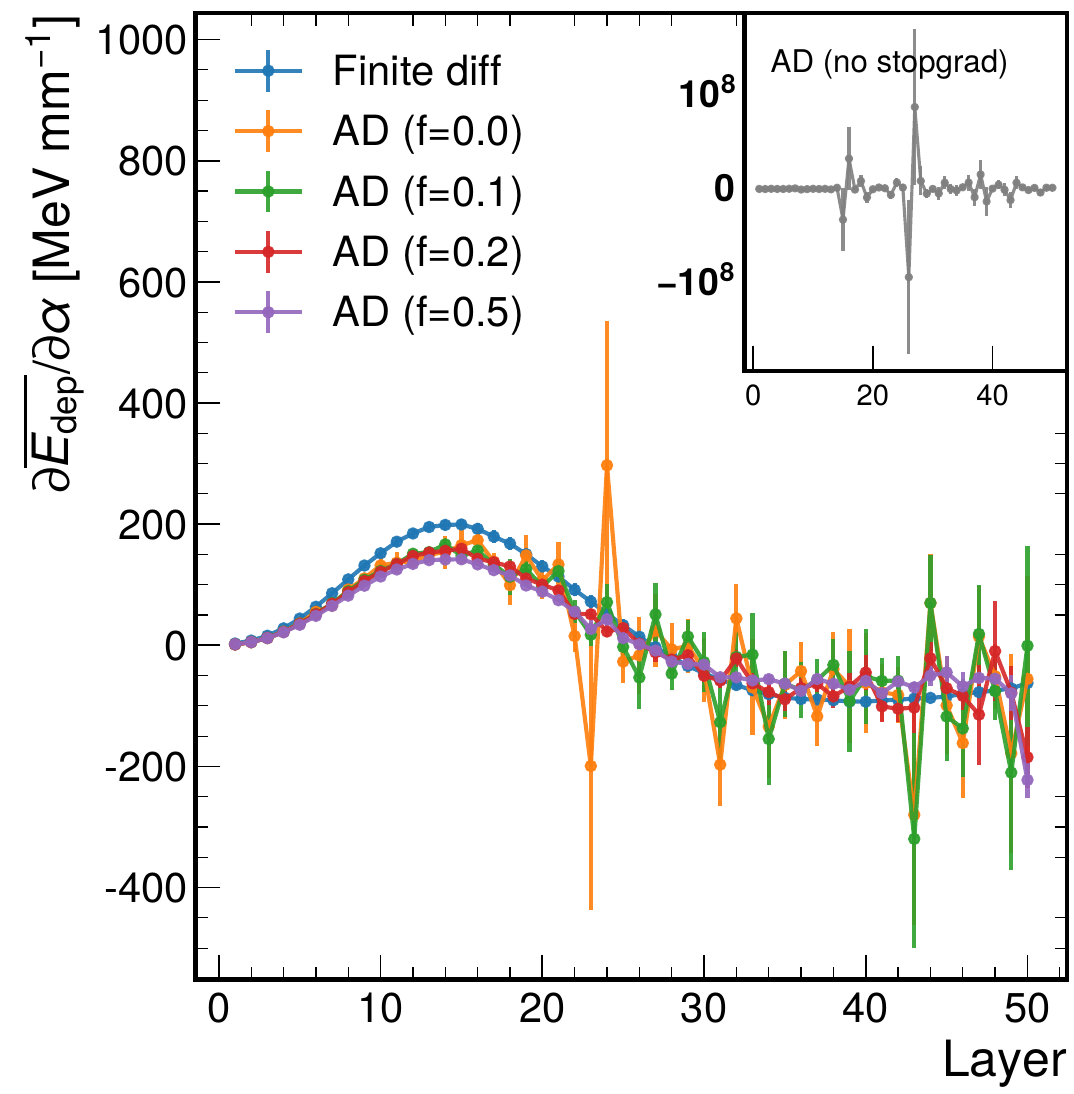}
    \caption{Mean longitudinal energy-deposit derivative per calorimeter layer comparing finite differences (blue) to AD with various choices of the stopgrad threshold $f$ (i.e.\ stopgrad is applied when $|\hat{v}\cdot\hat{n}|<f$).
    }
    \label{fig:hyperparameter-scans}
\end{figure}

To study the dependence on $f$, the stopgrad threshold $f$ is scanned for $f \in \{0.0, 0.1, 0.2, 0.5\}$.  
The results are shown in~\autoref{fig:hyperparameter-scans}. 
The central values of the AD derivatives are stable across this range with shapes generally close to the finite difference, most notably deviating by an underprediction of 25\% at layer 14.
Smaller values of $f$, such as $f=0$, demonstrate large variance at specific layers.
Larger values of $f$, such as $f=0.5$ suppress more derivative contributions from near-boundary steps, slightly reducing variance at the cost of increased bias.

Additionally, the photon MFP cap is scanned over $\lambda_{\mathrm{cap}} \in \{100, 1000, 10000\}$~mm, and the MSC conversion regularization parameter is scanned over $\epsilon_{\mathrm{conv}} \in \{10^{-4}, 10^{-3}, 10^{-2}\}$. 
These results are shown in~\autoref{sec:appendix} (\autoref{fig:regularization-scans}).
The choices $\lambda_{\mathrm{cap}} = 1000$~mm and $\epsilon_{\mathrm{conv}} = 10^{-3}$ exhibit minimal additional bias while providing further variance reduction. 
For these reasons, we adopt the working point $f = 0.2$, $\epsilon_{\mathrm{conv}} = 10^{-3}$, and $\lambda_{\mathrm{cap}} = 1000$~mm for the main text of this paper.
A cross-check on the robustness of the stopgrad strategy is presented in~\autoref{sec:appendix} (\autoref{fig:optimization_f0p05}), using a looser threshold of $f = 0.05$.
Good convergence of the optimization is obtained for both choices of $f$.

\subsection{Optimization}\label{sec:optimization}

\begin{figure}[h!]
    \centering
    \includegraphics[width=0.99\linewidth]{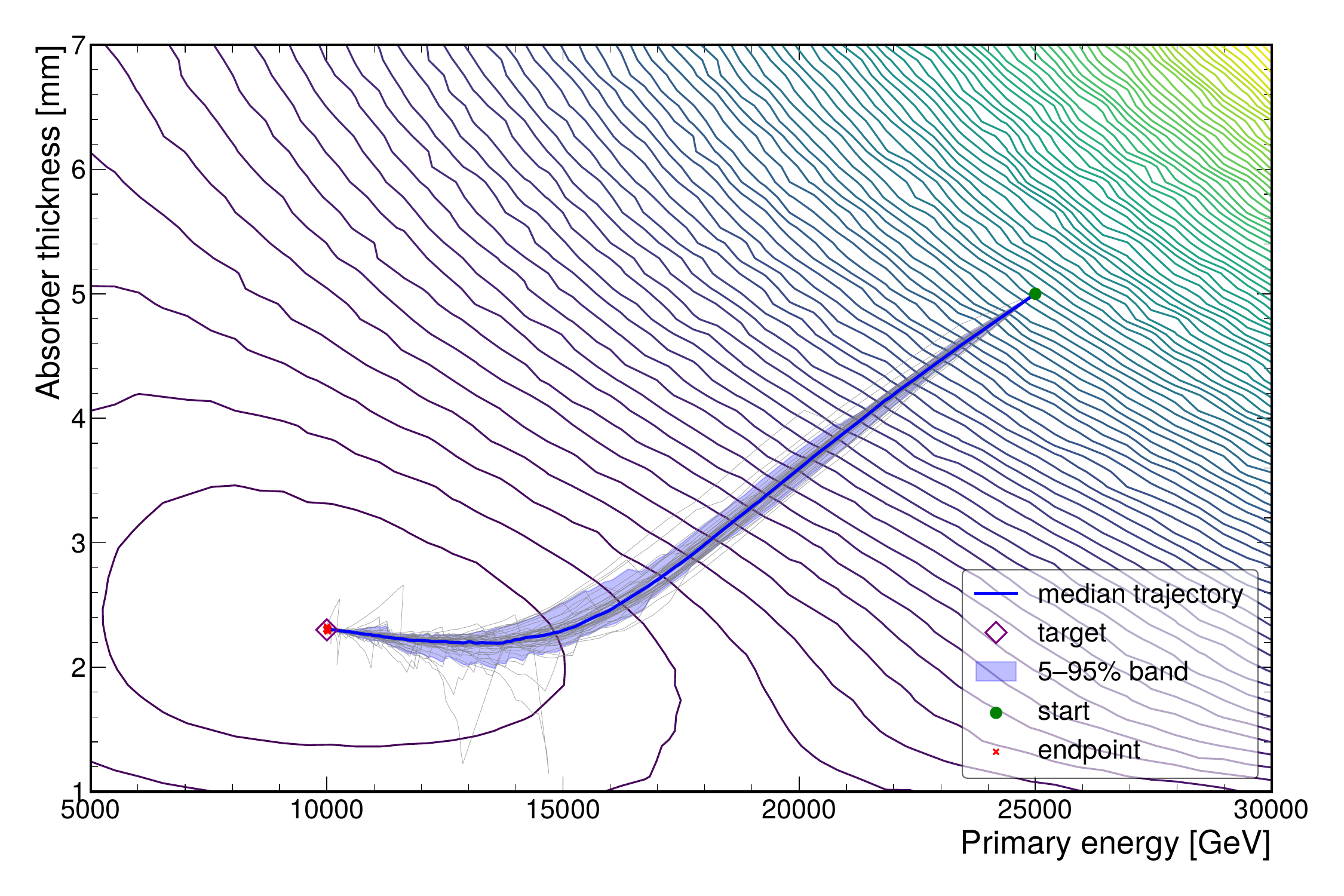}
    \caption{Trajectories of 50 randomly-seeded optimization runs minimizing the longitudinal-profile loss.
    Each grey curve shows one run, initialized at $(E_{\mathrm{beam}},\alpha)=(25000~\mathrm{MeV},\,5.0~\mathrm{mm})$ (green circle) and terminating at the final iterate (red cross).
    Contours show the loss landscape; the target $(E_{\mathrm{beam}}^*,\alpha^*)$ is marked with a purple diamond.
    The blue curve shows the median trajectory with a 5--95\% band.}
    \label{fig:optimization}
\end{figure}

We perform 50 optimization runs with identical hyperparameters and different random seeds. 
Their trajectories are shown in~\autoref{fig:optimization}.
All optimization runs descend into the basin of the minimum (2D contours) and terminate near the target parameters $(E_{\mathrm{beam}}^*,\alpha^*)$ (shown as a purple diamond). 
We summarize the paths of each optimization using the median trajectory (blue) with a 5--95\% band.

A stochastic gradient descent strategy is employed with a per-coordinate learning rate of $(1,\; 1\times10^{-7}~\text{mm}^2\,\text{MeV}^{-2})$, where the ratio is set by the natural scale $a^2/E_0^2$ to equalize the fractional step sizes across the two parameters.
Gradient clipping is found to be infrequent, triggered in only $0.02\%$ of gradient updates (2 out of $1\times10^4$ total steps across 50 independent seeds), indicating that the gradients remain stable throughout the optimization.
The final parameters of the optimizations are $E_\text{beam}^{\text{final}}= 9995 \pm6$~MeV, and $\alpha^{\text{final}}  = 2.307 \pm 0.007$~mm, consistent with $E_{\mathrm{beam}}^*$ and $\alpha^*$.

An optimization with identical settings, except using a looser threshold of $f=0.05$ is shown in~\autoref{sec:appendix} (\autoref{fig:optimization_f0p05}). Similar convergence quality with a larger fraction of clipped steps is observed.

\section{Conclusion}\label{conclusion}

We studied automatic differentiation in a \textsc{Geant4}-like electromagnetic radiation transport simulation and identified rare, geometry-coupled code paths that produce exploding gradient variance, making naïve derivatives unusable for optimization. 
We observed that large-variance gradients were strongly correlated with a small subset of boundary-limited transport configurations, particularly those involving near-parallel incidence and backward-going intersections on boundaries, which can trigger repeated boundary interactions and amplify geometry-induced sensitivities.

To mitigate these instabilities, we introduced a targeted navigation stop-gradient rule applied only to geometry-induced updates during boundary-limited stepping under simple kinematic conditions. 
This intervention leaves the forward (primal) simulation unchanged while suppressing catastrophic outliers by 6 orders of magnitude, yielding stable, optimization-ready derivatives with a complete electromagnetic physics model including multiple Coulomb scattering. 

Finally, we demonstrated the utility of the derivatives in a detector optimization, where ensembles of randomly seeded runs reliably converged to the target shower-profile parameters. 
These results establish practical criteria for numerical stability in differentiable particle transport.
Furthermore, they suggest a path toward gradient-based detector and radiation-system design in high-dimensional settings, with potential applications including medical physics, shielding, and space instrumentation.

Ultimately, differentiable simulation has the potential to close the loop between detector design, simulation, and analysis, enabling end-to-end optimization pipelines that were previously intractable. 
Scaling to full hadronic physics and realistic detector geometries remain the key challenges, but the stability criteria identified here provide a foundation for systematically tackling these more complex settings.

\section*{Acknowledgements}
J.K. and M.K. are supported by the US Department of Energy (DOE) under Grant No. DE-AC02-76SF00515.
The authors from RPTU University Kaiserslautern-Landau appreciate funding from the German National High Performance Computing (NHR) association for the Center NHR South-West. 
L.H. is supported by BMFTR Project SciFM 05D2025 and by the Excellence Cluster ORIGINS, which is funded by the Deutsche Forschungsgemeinschaft (DFG, German Research Foundation) under Germany’s Excellence Strategy - EXC-2094-390783311.
D.L. and V.V. are supported by the National Science Foundation (NSF) under Grant OAC-2311471.

\section*{Code availability}
The simulation code and derivative-regularization implementations used in this work are publicly available.
The modified \textsc{g4hepem} physics engine is at \url{https://github.com/jeffkrupa/g4hepem/releases/tag/v1.0-paper}, and the \textsc{hepemshow} application is at \url{https://github.com/jeffkrupa/hepemshow/releases/tag/v1.0-paper}.

\bibliographystyle{iopart-num}

\bibliography{bibliography}

\appendix

\section{Supplementary Results}\label{sec:appendix}

\begin{table}[htbp]
\centering
\begin{tabular}{@{}cccccccc@{}}
\toprule
\multirow{2}{*}{$f$} & \multirow{2}{*}{Propagation} & \multicolumn{3}{c}{Stopgrad Fraction (\%)} \\
\cmidrule(lr){3-5}
& & Energy & Steps & Tracks \\
\midrule
0.0  & Track-only & 9.3  & 20.2 & 11.8 \\
     & Track + desc. & 15.1 & 26.9 & 32.3 \\
\midrule
0.05 & Track-only & 9.6  & 20.4 & 11.9 \\
     & Track + desc. & 15.5 & 27.2 & 32.8 \\
\midrule
0.1  & Track-only & 9.9  & 20.7 & 12.2 \\
     & Track + desc. & 16.0 & 27.8 & 33.6 \\
\midrule
0.2  & Track-only & 10.8 & 21.6 & 12.9 \\
     & Track + desc. & 17.6 & 29.3 & 36.0 \\
\midrule
0.5  & Track-only & 16.0 & 27.2 & 16.3 \\
     & Track + desc. & 25.9 & 37.5 & 46.7 \\
\bottomrule
\end{tabular}
\caption{From left to right, the columns show the percentages of energy deposited by stopgrad steps, stopgrad steps, and tracks containing a stopgrad step.
\emph{Track-only} applies stopgrad to the remainder of the affected track; \emph{Track + desc.}\ additionally propagates stopgrad to all descendant particles.}
\label{tab:stopgrad_fractions}
\end{table}

\begin{figure}
    \centering
    \includegraphics[width=0.47\linewidth]{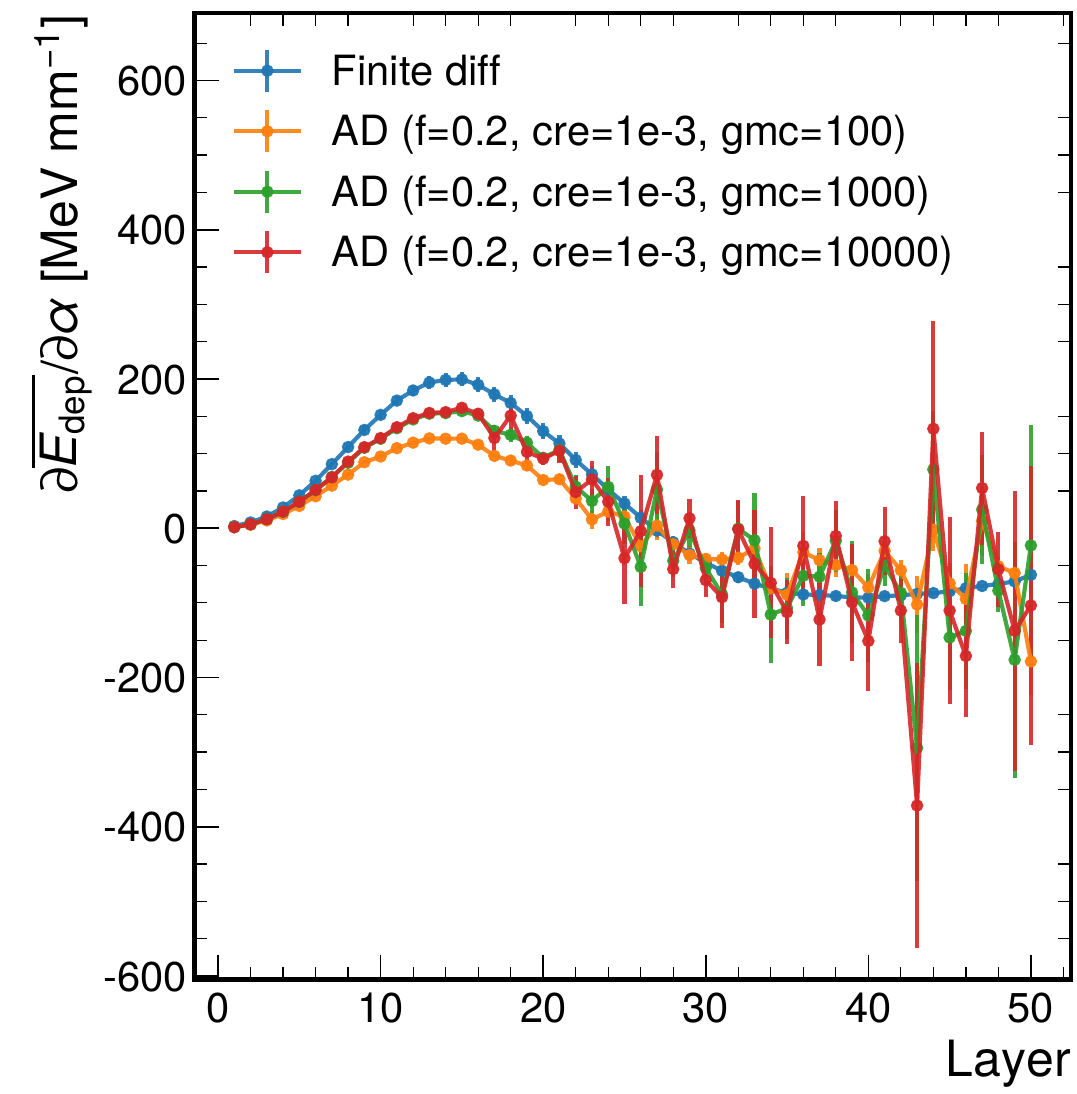}
    \includegraphics[width=0.48\linewidth]{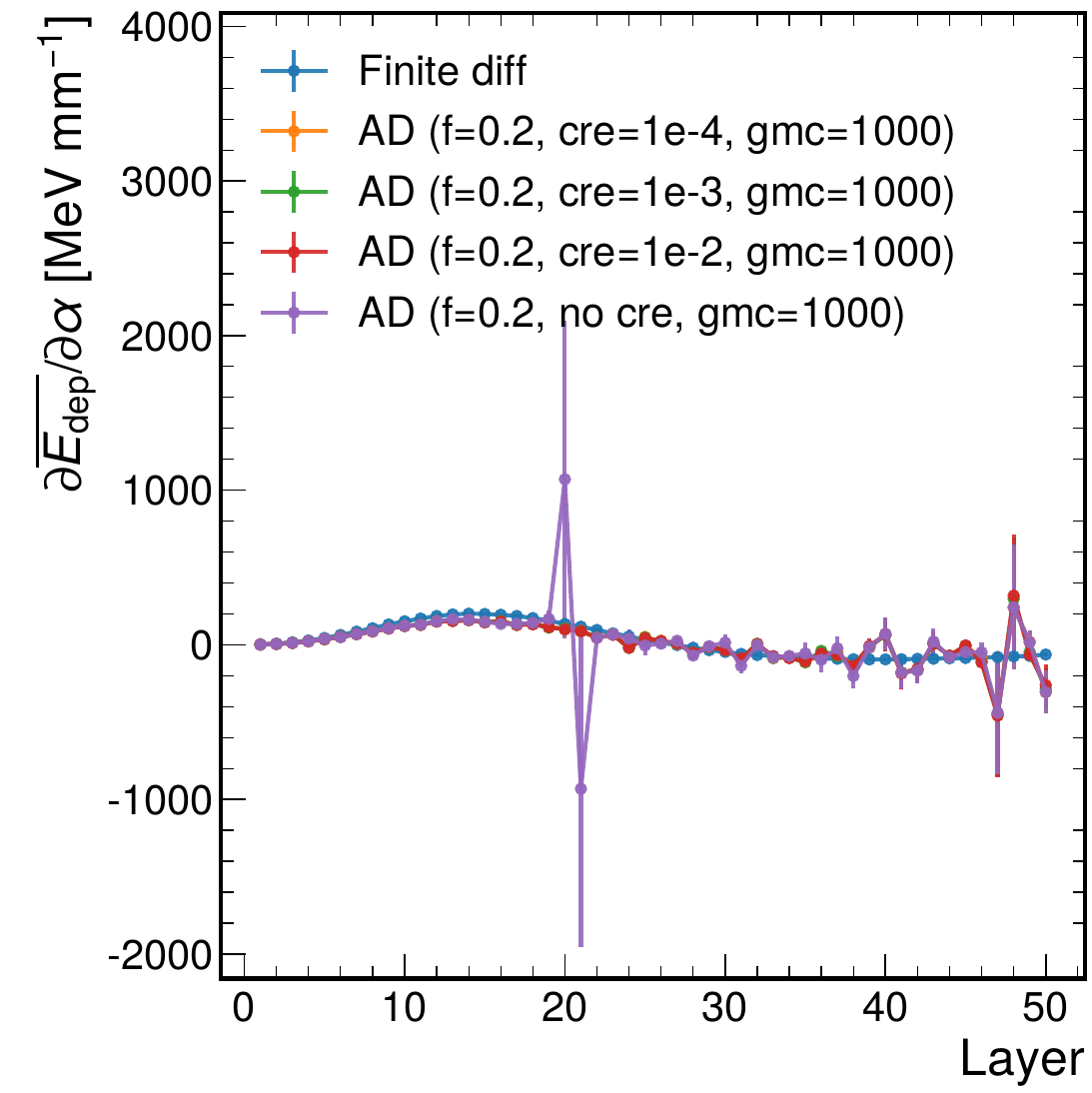}
    \caption{Mean longitudinal energy-deposit derivative per calorimeter layer comparing finite differences (blue) to AD with various settings. 
    Left: a scan over parameter $\lambda_\text{cap}$.
    Right: a scan over parameter $\epsilon_\text{conv}$.
    }
    \label{fig:regularization-scans}
\end{figure}

\begin{figure}
    \centering
    \includegraphics[width=0.99\linewidth]{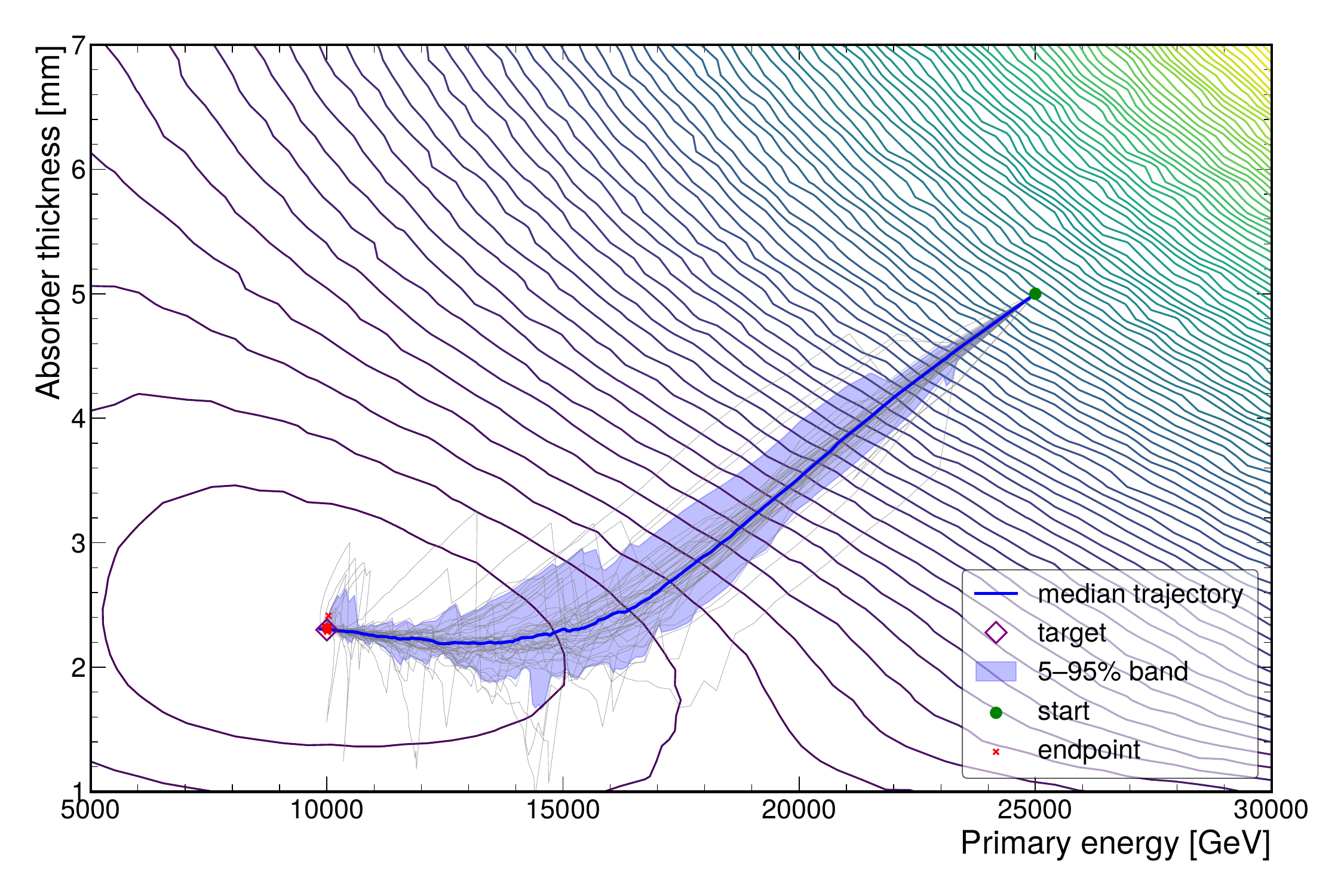}
    \caption{Trajectories of 50 randomly-seeded optimization runs minimizing the longitudinal-profile loss.
    Each grey curve shows one run, initialized at $(E_{\mathrm{beam}},\alpha)=(25000~\mathrm{MeV},\,5.0~\mathrm{mm})$ (green circle) and terminating at the final iterate (red cross).
    Contours show the loss landscape; the target $(E_{\mathrm{beam}}^*,\alpha^*)$ is marked with a purple diamond.
    A regularization of $f=0.05$, $\epsilon_{\mathrm{conv}} = 10^{-3}$, and $\lambda_{\text{cap}}=1000$~mm is used. 
    The blue curve shows the median trajectory with a 5--95\% band.}
    \label{fig:optimization_f0p05}
\end{figure}

\end{document}